\documentclass[a4paper,fleqn,usenatbib]{mnras}

\usepackage[T1]{fontenc}
\usepackage{ae,aecompl}

\usepackage{graphicx}	
\usepackage{amsmath}	
\usepackage{amssymb}	
\usepackage{multirow}
\usepackage{txfonts}
\usepackage[utf8]{inputenc}
\usepackage{float}
\usepackage{subcaption}

\usepackage{subfiles} 

\usepackage[decimalsymbol=., expproduct=times, separate-uncertainty=true, multi-part-units=single]{siunitx} 

\title[Viscous evolution of disks in star clusters]{The viscous evolution of circumstellar discs in young star clusters}

\author[Concha-Ramírez et al.]{
Francisca Concha-Ramírez\thanks{E-mail: fconcha@strw.leidenuniv.nl},
Eero Vaher,
Simon Portegies Zwart
\\
Leiden Observatory, Leiden University, PO Box 9513, 2300 RA Leiden, The Netherlands\\
}

\date{Accepted XXX. Received YYY; in original form ZZZ}

\pubyear{2018}

\begin{document}

\label{firstpage}
\pagerange{\pageref{firstpage}--\pageref{lastpage}}
\maketitle

\begin{abstract}
Stars with circumstellar disks may form in environments with high stellar and gas densities which affects the disks through processes like truncation from dynamical encounters, ram pressure stripping, and external photoevaporation. Circumstellar disks also undergo viscous evolution which leads to disk expansion. Previous work indicates that dynamical truncation and viscous evolution play a major role in determining circumstellar disk size and mass distributions. However, it remains unclear under what circumstances each of these two processes dominates. Here we present results of simulations of young stellar clusters taking viscous evolution and dynamical truncations into account. We model the embedded phase of the clusters by adding leftover gas as a background potential which can be present through the whole evolution of the cluster, or expelled after $\SI{1}{Myr}$. We compare our simulation results to actual observations of disk sizes, disk masses, and accretion rates in star forming regions. We argue that the relative importance of dynamical truncations and the viscous evolution of the disks changes with time and cluster density. Viscous evolution causes the importance of dynamical encounters to increase in time, but the encounters cease soon after the expulsion of the leftover gas. For the clusters simulated in this work, viscous growth dominates the evolution of the disks.
\end{abstract}


\section{Introduction}

Stars are formed in clustered environments \citep{Clarke2000, Lada2003}. Circumstellar disks develop shortly after star formation \citep{2011ARA&A..49...67W}, when they are still embedded in the dense star and gas surroundings of the cluster. In these environments, disks can be disturbed by different processes such as truncations due to close stellar encounters \citep{Rosotti2014, 2015A&A...577A.115V, PortegiesZwart2016}, external photoevaporation due to nearby bright O type stars \citep{Odell1998, Scally2001, Guarcello2016, 2017arXiv170303409H}, accretion and ram pressure stripping \citep{Wijnen2016, 2017arXiv170204383W}, and nearby supernovae \citep{2012MNRAS.420.1503P}. Circumstellar disks can also present viscous evolution \citep{Lynden-Bell1974}. As the typical viscous time scales of T~Tauri stars seem to be on the order of \SI{e5}{yr} \citep{Hartmann1998,Isella2009}, circumstellar disks are likely to undergo considerable viscous growth during the first few million years of cluster evolution.

Studying the processes that affect the distribution of circumstellar disks in young star clusters helps to understand the development of planetary systems like our own. Different processes, however, can dominate the evolution of clusters at different times and cluster densities. Young clusters are still embedded in the gas from which they formed. Gas expulsion, which can be the result of feedback from massive stars such as winds and supernovae explosions, leaves the cluster in a supervirial state that leads to its expansion and possible dissolution \citep{tutukov1978}. \citet{Vincke2016} carried out simulations including the effect of truncations by stellar encounters before and after gas expulsion. They show that taking the early gas expulsion into account in their simulations increases the rate of stellar encounters, because the larger total mass increases the stellar velocity dispersion.

Most approaches to study the interaction of the circumstellar disks with their surrounding cluster have focused on a static size for the disk which can only decrease by the influencing external processes (e.g., \citet{Scally2001,Pfalzner2006,Olczak2006,Olczak2010,Vincke2015,PortegiesZwart2016}). In contrast, \citet{Rosotti2014} considered the viscous evolution of the disks along with truncations by dynamical encounters, by combining N-body simulations with smoothed particle hydrodynamics (SPH) to represent the growth of the disks. However, due to the numerical cost of their simulation, their study was limited to 100 stars of $1 M_{\odot}$ each distributed in a Plummer sphere, only half the stars had a circumstellar disk, and their simulations were run for just $\SI{0.5}{Myr}$.

The purpose of this work is to analyze the combined effect of viscous disk evolution and the presence of gas on the dynamics and circumstellar disk distributions in young star clusters. We look to understand the relative importance of such processes during different stages of cluster evolution. We include the viscous evolution of the circumstellar disks semi-analytically using the similarity solutions developed by \citet{Lynden-Bell1974}. We also consider disk truncations caused by dynamical encounters between stars in the cluster. Aditionally, we model the presence of gas in the cluster, as a means to represent the embedded phase, and the further expulsion of said gas. We carry out our simulations using the AMUSE framework \citep{2013CoPhC.183..456P, 2013A&A...557A..84P}. All the code used for the simulations, data analyses, and figures of this paper is available in a Github repository\footnote{\url{http://doi.org/10.5281/zenodo.1465931}}.

Thanks to modern observational techniques, circumstellar disks have been observed and characterized inside many open star clusters and star forming regions, such as Chamaeleon I \citep{2016ApJ...831..125P, 2017ApJ...847...31M, 2017arXiv170402842M}, $\sigma$ Orionis \citep{2016ApJ...829...38M, 2017AJ....153..240A}, the Lupus clouds \citep{2018ApJ...859...21A, 2016ApJ...828...46A, 2014A&A...561A...2A}, the Orion Trapezium cluster \citep{Vicente2005, 2009ApJ...694L..36M, 2004ApJ...606..952R}, and the Upper Scorpio region \citep{2017ApJ...851...85B, 2016ApJ...827..142B}. From these observations, the size and mass of the disks and the accretion rate onto their central star can be calculated. This brings an opportunity to calibrate the results obtained by simulations, by offering a way to compare simulated disk distributions with observed ones.

\section{Methods}
\label{sec:methods}

\subsection{Evolution of isolated viscous disks}
\label{subsec:isolated_disk}

\citet{Lynden-Bell1974} showed that for a thin disk in which viscosity has a radial power-law dependence and no time dependence, there exists a similarity solution to which all initial mass distributions will asymptotically approach. The description of the similarity solutions used in this work is largely based on the one provided by \citet{Hartmann1998}, who applied the similarity solutions to explain the observed accretion rates of T~Tauri stars.

The similarity solutions of viscous disks are characterised by four independent parameters:
\begin{enumerate}
    \item $\gamma$ - the radial viscosity dependence exponent.
    \item $M_d(0)$ - the initial disk mass.
    \item $R_c(0)$ - the initial characteristic disk radius, outside of which $1/e\simeq\SI{37}{\percent}$ of the disk mass initially resides.
    \item $t_v$ - the viscous time scale at $R_c(0)$. It is possible to instead specify $v_c$, the viscosity at $R_c(0)$.
\end{enumerate}

For the present work, the modeled disks will be represented by three properties: their characteristic radius, the mass of the disk, and the accretion rate from the disk to the central star. These are the properties derived from observations, allowing us to compare them directly with the simulation results. 

According to the model developed by \citet{Lynden-Bell1974}, the characteristic radius of the disk as a function of time $t$ is given by
\begin{equation}\label{eq:diskradius}
    R_c(t)=\left(1+\frac{t}{t_v}\right)^\frac{1}{2-\gamma}R_c(0).
\end{equation}

\noindent
The disk mass as a function of time is
\begin{equation}
    M_d(t)=M_d(0)\left(1+\frac{t}{t_v}\right)^\frac{1}{2\gamma-4}
    \label{eq:mdisk}
\end{equation}

\noindent
and the radial cumulative mass distribution of the disk as a function of time is
\begin{equation}
    M_d(R,t)=M_d(t)\left(1-e^{-\Gamma}\right).
    \label{eq:mdensity}
\end{equation}
Here
\begin{equation}\label{eq:gamma}
\Gamma = \left(\frac{R}{R_c(0)}\right)^{2-\gamma} \left(1+\frac{t}{t_v}\right)^{-1}
\end{equation}

The accretion rate of matter to the star as a function of time is 
\begin{equation}
    \dot{M}_{acc}(t)=-\frac{\text{d}M_d(t)}{\text{d}t}=\frac{M_d(0)}{(4-2\gamma)t_v}\left(1+\frac{t}{t_v}\right)^\Delta.
    \label{eq:accretion_rate}
\end{equation}
Here $\Delta = \frac{5-2\gamma}{2\gamma-4}$. The viscous time scale $t_v$ relates to the characteristic viscosity of the disk $v_c$ by
\begin{equation}
    t_v=\frac{R_c(0)^2}{3(2-\gamma)^2v_c}.
    \label{eq:tv_from_vc}
\end{equation}
The disk viscosity can be written as
\begin{equation}
    v(R)=\alpha \frac{c_s^2}{\Omega}=\alpha \frac{k_BT\sqrt{R^3}}{\mu m_p\sqrt{GM_*}}.
\end{equation}
Here $\alpha$ is the turbulent mixing strength \citep{Shakura1973}, $c_s=\sqrt{\frac{k_BT}{\mu m_p}}$ is the isothermal sound speed, $\Omega=\sqrt{\frac{GM_*}{R^3}}$ is the Kepler frequency, $k_B$ is the Boltzmann constant, $T$ is the disk temperature at distance $R$ from the star, $\mu$ is the mean molecular weight of the gas in atomic mass units, $m_p$ is the proton mass, $G$ is the constant of gravity, $M_*$ is the mass of the central star and the mass of the disk is assumed to be insignificant. Considering the temperature of the disk as a radial power law $T\propto R^{-q}$,
\begin{equation}
    v_c=\alpha\frac{k_BTR_c(0)^{\frac{3}{2}-q}}{\mu m_pR^{-q}\sqrt{GM_*}},
    \label{eq:viscosity}
\end{equation}
and
\begin{equation}
    t_v=\frac{\mu m_pR_c(0)^{\frac{1}{2}+q}\sqrt{GM_*}}{3\alpha\left(2-\gamma\right)^2k_B T R^{q}}.
     \label{eq:viscous_timescale}
\end{equation}

Relating $T_d$ with an estimate of stellar luminosity $L_* = L_*(M_*)$ allows us to write $t_v = t_v(M_*, \alpha, \gamma, q, R_c(0))$. While $t_v$ is an independent parameter for the similarity solutions, we parametrize it as a function of $\gamma$ and $R_c(0)$ in our implementation, on physical grounds. We use the zero age main sequence stellar luminosities for stars with metallicities Z = 0.02 from \citet{Hurley2000} that are available in the package SeBa, which is incorporated into AMUSE.

If it is observed that $\dot{M}_{acc}(t) \propto t^{-\eta}$, then the viscosity exponent $\gamma$ can be defined as:

\begin{equation}
    \gamma=\frac{4\eta-5}{2\eta-2},
\end{equation}
where we assumed that $\gamma$ is the same for all disks. Previous work by \citet{Hartmann1998}, \citet{SiciliaAguilar2010} and \citet{Antoniucci2014} have found the value of $\eta$ to be in the range $\num{1.2}\lesssim\eta\lesssim\num{2.8}$. Furthermore, \citet{Andrews2010} found a value of $\gamma = 0.9 \pm 0.2$ in the Ophiuchus star forming region, independent of the disk masses or stellar properties. In general $\gamma=1$, which corresponds to $\eta=\num{1.5}$, seems to be in agreement with most observed disks and is thus the value we adopt in this work.

\begin{figure}
    \includegraphics[width=\columnwidth]{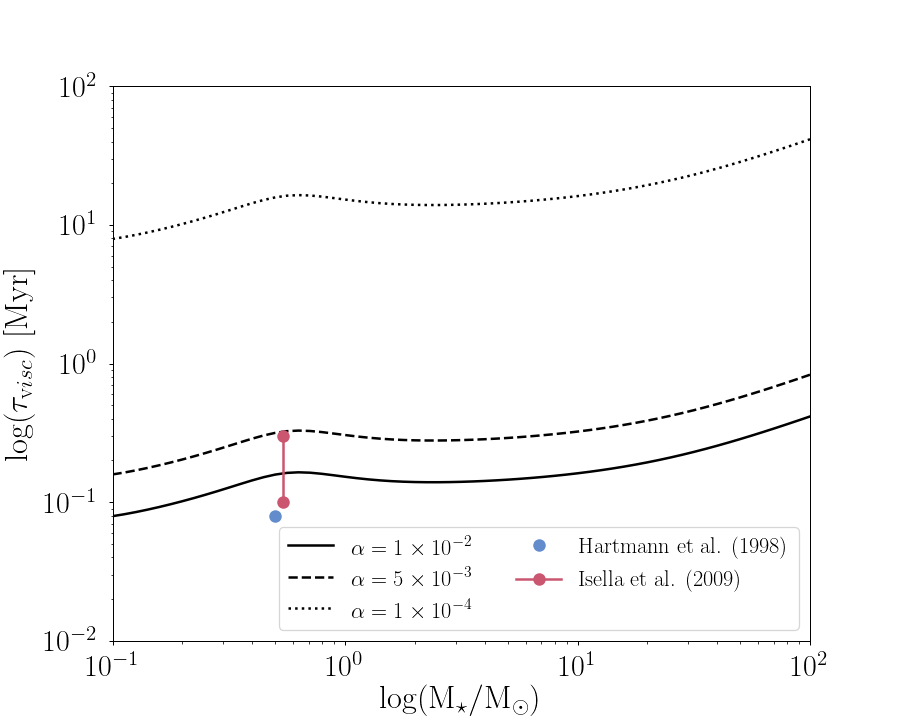}
    \caption{Viscous time scales in our model as a function of stellar mass, for different values of the turbulence parameter $\alpha$. Observational estimates for circumstellar disks around T~Tauri stars by \citet{Hartmann1998} and \citet{Isella2009} are shown for comparison.}
    \label{fig:viscous_timescales}
\end{figure}

Equation~(\ref{eq:viscous_timescale}) shows that the viscous time scale depends on the turbulence parameter $\alpha$ and on the temperature profile of the disk. A value of $\alpha\sim\num{e-2}$ was found by \citet{Hartmann1998} from observations of T~Tauri stars. \citet{Isella2009} suggest that $\alpha$ might range from \numrange{e-4}{0.5}, while \citet{Andrews2010} found $\alpha\sim\numrange[range-phrase = -]{e-3}{e-2}$ in the Ophiuchus star forming region. \citet{Mulders2012} found that $\alpha\sim\num{e-4}$ for circumstellar disks around stars of all masses, but by assuming slightly different circumstellar dust properties $\alpha\sim\num{e-2}$ could also fit the observations. 
In figure~\ref{fig:viscous_timescales} we show the viscous time scales obtained in our model for three different values of the turbulence parameter: $\alpha=\num{e-4}$, $\alpha=\num{5e-3}$, and $\alpha=\num{e-2}$. It can be seen that the two highest values of the turbulence parameter lead to viscous time scales that agree with observations, up to moderately massive stars. Motivated by this, we adopt two values for this parameter: $\alpha=\num{e-2}$, in what we refer to as fast viscous evolution, and $\alpha=\num{5e-3}$ in what we call slow viscous evolution. The viscosity parameters chosen are in good agreement with estimates for the viscous time scale, $t_v$. \citet{Hartmann1998} found a value of $t_v\sim\SI{8e4}{yr}$ for a
typical T~Tauri star. \citet{Isella2009} estimate, based on their
observations, that $t_v\sim\SIrange[range-phrase =
  -]{e5}{3e5}{yr}$. 

Values of $\alpha$ should be considered only an approximation. In reality, disk parameters related to the viscosity ---like the viscosity exponent $\gamma$ and the turbulence parameter $\alpha$--- should be expected to vary among stars of equal mass, age, and observed accretion rates; even within the same disk, different parts of it may show different viscosity parameters \citep{2016ApJ...816...25P}.

\subsection{Gas in the cluster}
\label{gas_presence}
The presence of gas in the cluster is modelled semi-analytically with a background potential in the $N$-body computations. We adopt the gas and gas expulsion parameters from \citet{Lughausen2012}. The distribution of the gas corresponds to a Plummer sphere with the same Plummer radius as the stars in the cluster. The initial mass of the gas, $M_g(0)$, is taken to be twice the mass of the stars, which corresponds to a star formation efficiency of 1/3.

In the runs with gas expulsion, the mass of the gas as a function of time is given by

\begin{equation}\label{eq:gas_exp_onset}
    M_g(t)=\begin{cases}
        M_g(0)&t\leq t_D,\\
        \frac{M_g(0)}{\varphi}&t>t_D,
    \end{cases}
\end{equation}
with $\varphi=1+(t-t_D)/\tau$, and where $t_D$ is the time at which gas expulsion begins. The gas expulsion timescale corresponds to the time it takes for the cluster to lose half its gas once gas expulsion has begun, and it is given by

\begin{equation}
\tau=\frac{R_P}{\SI{1}{pc}}\times\SI{0.1}{Myr},
\end{equation}
where $R_P$ is the Plummer radius of the cluster.

\subsection{Dynamical disk truncation}
\label{subsec:truncation_procedure}

The semi-analytical model used in this work assumes that, between encounters, disks evolve according to the similarity solution described in section \ref{subsec:isolated_disk}. A close enough stellar encounter introduces a discontinuity in the disk parameters, but after that the disk is assumed to continue evolving as an isolated viscous disk.

The work by \citet{Rosotti2014} encourages us to adopt the commonly used approximation that an encounter truncates circumstellar disks at one third of the encounter distance in the case of equal massed stars. In addition, we use the mass dependence from \citet{Bhandare2016}. 
The combination of these two models gives us the characteristic radius of a circumstellar disk immediately after an truncating encounter:
\begin{equation}
    R'_c=\frac{r_\text{enc}}{3}\left(\frac{M}{m}\right)^{\num{0.2}},
    \label{eq:truncation_radius}
\end{equation}
where $r_\text{enc}$ is the encounter distance, $M$ is the mass of the star with the disk in question, and $m$ is the mass of the encountered star.

Equation~(\ref{eq:mdisk}) gives the disk mass immediately before the encounter. This mass is lower than the initial disk mass, because some of it has been accreted onto the star. In our model this mass difference is added to the stellar mass. If the encounter is assumed to not remove mass from the inner part of the disk, then the disk mass inside the new initial characteristic radius just after the encounter can be assumed to be equal to the mass inside that radius just before the encounter. Equation~(\ref{eq:mdensity}) and the properties of characteristic radius can then be used to find the disk mass immediately after the encounter as
\begin{equation}
    M'_d=\frac{M_d(R'_{c},t)}{1-\frac{1}{e}}\simeq\num{1.6}M_d(R'_{c},t),
\end{equation}
where $t$ is the time from the last discontinuity of the disk parameters.

According to equation~(\ref{eq:tv_from_vc}) and the underlying assumption $v\propto R^\gamma$, the new viscous timescale of the disk will be
\begin{equation}
t'_v=\left(\frac{R'_c}{R_c(0)}\right)^{2-\gamma}t_v.
\label{eq:visc}
\end{equation}

In our model, the relative change in disk mass and accretion rate in an encounter is determined by $\gamma$ and the truncation radius relative to the disk characteristic radius just before the encounter, $R'_c/R_c(t)$. If the disk is truncated, its viscous time scale decreases and the disk starts evolving faster. We ignore the orientation of the disks, so the equation for truncation radius is the average truncation radius over all disk inclinations.

\subsection{Numerical implementation}
\label{subsec:numerical}

We use the 4th-order Hermite $N$-body code {\tt ph4}, which is incorporated into the AMUSE framework. The softening length is $\epsilon=\SI{100}{au}$ and the timestep parameter is $\eta=\num{e-2}$. The simulations start in virial equilibrium and last for \SI{2}{Myr}. 

Collisional radii for each star in the cluster are defined depending on the stellar mass and the theoretical size of the disk evolving in isolation, given by equation \ref{eq:diskradius}. The initial collisional radius for each star is given by the distance at which the most massive star in the cluster can truncate its disk. The collisional radii of all stars are updated every \SI{2000}{yr}, to account for the viscous evolution of the disks. Two stars are considered to be in an encounter if the distance between them is less than the sum of their collisional radii. The encounter distance between the stars is then computed by analytically solving the two body problem. If the encounter is strong enough as to truncate the disks, disk parameters of both stars are updated as described in section \ref{subsec:truncation_procedure}. After each encounter, the collisional radii of both stars are set to $\num{0.49}r_\text{enc}$ in order to prevent the $N$-body code from detecting the same encounter multiple times during the same \SI{2000}{yr} window.

\section{Results}
\label{sec:results}

\subsection{Initial conditions}\label{subsec:ics}

\subsubsection{Cluster properties}

Simulations were run for clusters with 1500 stars. The stellar masses
are randomly sampled from a Kroupa initial mass distribution
\citep{Kroupa2001} with lower mass limit $0.1 M_\odot$ and upper mass
limit $100 M_\odot$. The stars are initially distributed in a Plummer
sphere \citep{Plummer1911} with Plummer radius 0.5 parsec. All stars
are assumed to be single and coeval.

All simulations were carried out for $\SI{2.0}{Myr}$. We define three different gas scenarios for the simulations:

\begin{enumerate}
\item No gas
\item Gas presence
\item Gas expulsion starting at $\SI{1.0}{Myr}$
\end{enumerate}

Each simulation was run for two values of the turbulence parameter: $\alpha=\num{e-2}$ and $\alpha=\num{5e-3}$.

\subsubsection{Disk properties}
Based on the observations of low mass stars carried out by \citet{Isella2009} and the estimations obtained by \citet{Hartmann1998}, the initial characteristic radius $R_c$ of the circumstellar disks was chosen to be

\begin{equation}
    R_c(0)=R'\left(\frac{M_*}{M_{\sun}}\right)^{0.5}
\end{equation}
with $R'=\SI{30}{AU}$. 

Circumstellar disks with masses larger than 10\% of the mass of their star are the most likely to be gravitationally unstable \citep{Kratter2016}. According to hydrodynamical simulations of collapsing gas clouds, the disk-to-star mass ratios of embedded protostars are large enough for gravitational instabilities to occur \citep{Vorobyov2011}, but these instabilities lead to accretion bursts that quickly decrease the mass ratios \citep{Armitage2001,Kratter2016}. 
Based on this we chose the initial disk masses to be 
\begin{equation}
M_d(0)=\num{0.1}M_*
\end{equation}

\subsection{The effect of gas in the cluster}\label{subsec:results_gas}

\begin{figure}
    \includegraphics[width=\columnwidth]{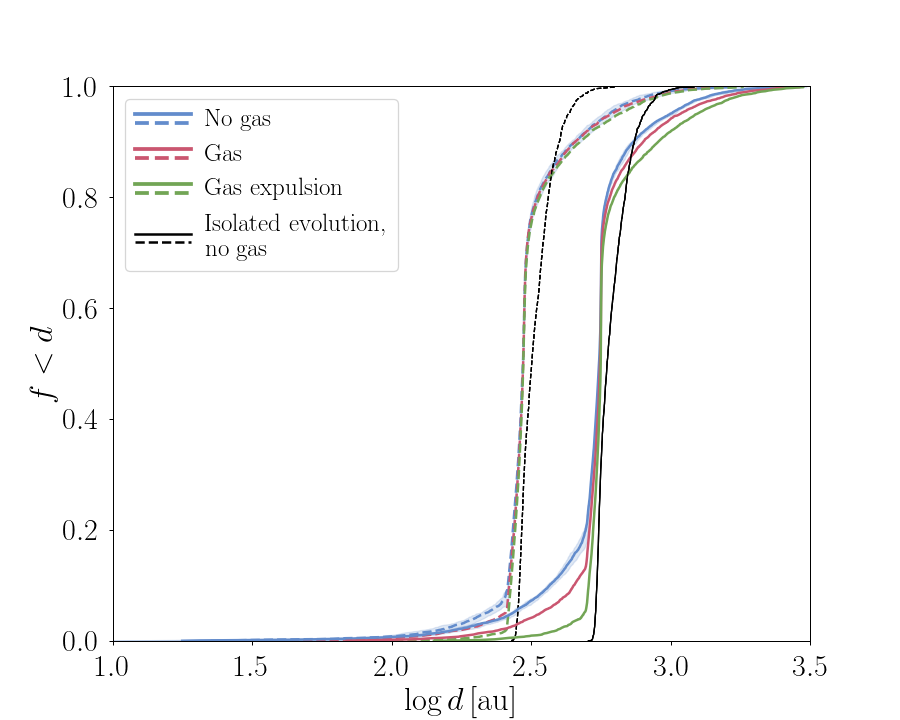}
    \includegraphics[width=\columnwidth]{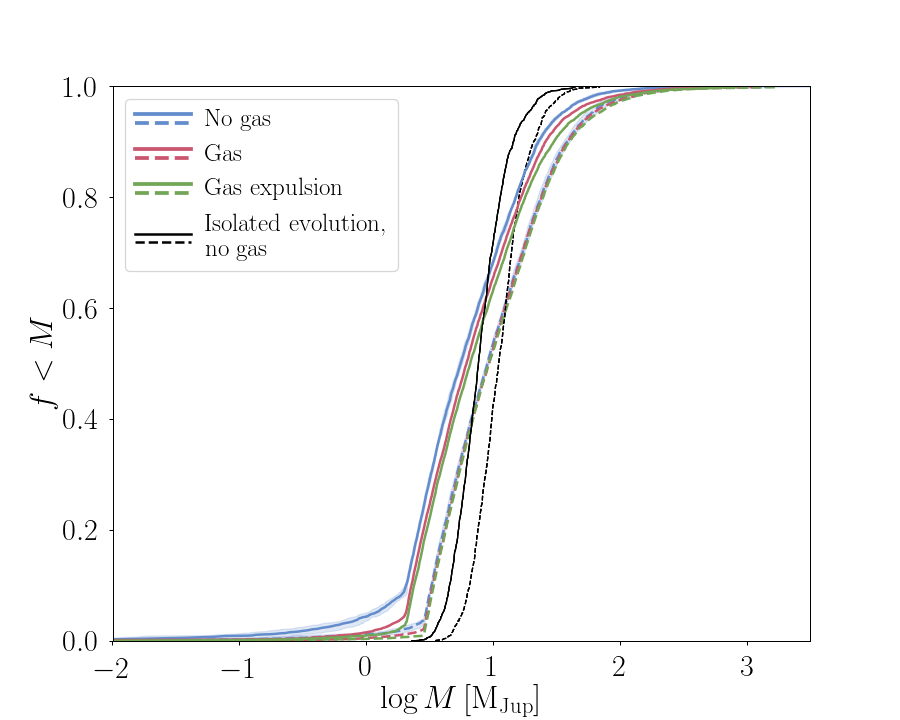}
    \includegraphics[width=\columnwidth]{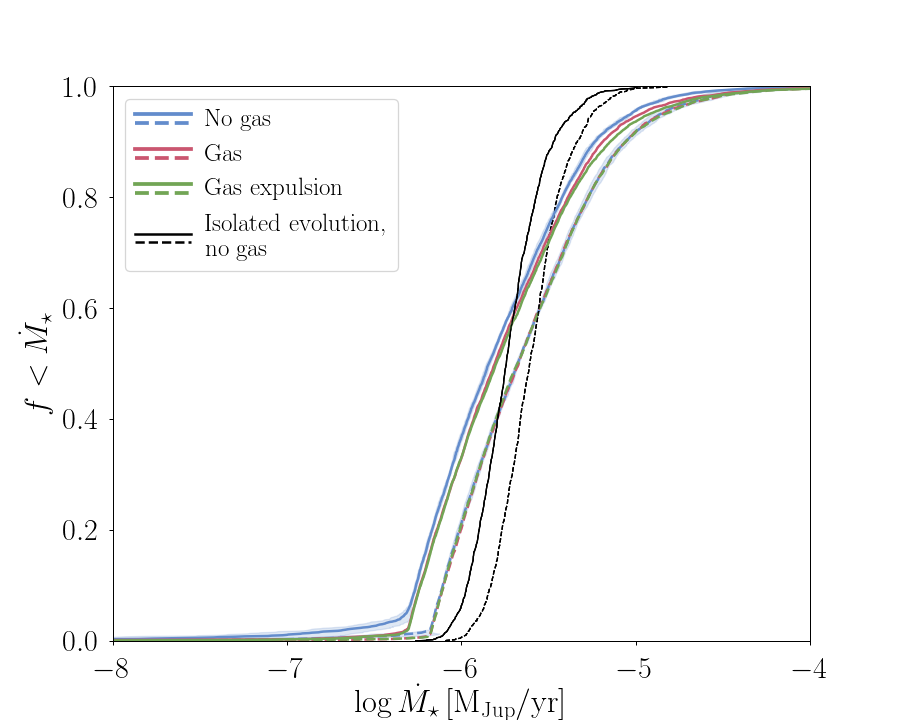}
    \caption{Cumulative distributions of the mean disk size (top
      panel) and mass (center), and the accretion rate onto central
      star (bottom panel) at the end of the simulations. The colors indicate the different mode of gas loss in
      the simulation (see legend in the top left corner), and in black we show the results of isolated disk evolution (viscous growth only, no truncations). The solid lines correspond to fast viscous disk evolution ($\alpha=\num{e-2}$) and the dashed lines to slow viscous evolution ($\alpha=\num{5e-3}$). The shaded areas around the blue curves indicate the dispersion around the mean value averaged over 5 simulations. For clarity we only show this uncertainty interval for the blue curves, but the others are comparable.}
    \label{fig:CDF}
\end{figure}

In Fig.\,\ref{fig:CDF} we present the cumulative distributions for the
mean values of sizes and masses of the circumstellar disks, and the
accretion rates onto the central star.  The three different colors in each
of the panels present one of our choices of how the gas is removed
from the cluster. In black we show the results of isolated evolution, where disks are subject to viscous growth but no dynamical truncations.
In addition, we present the results for two choices
of the turbulence parameter $\alpha$ with solid and dashed lines.

Changing the value of $\alpha$ has quite a pronounced effect on the
size distribution of the disks, in the sense that a low value (of
$\num{5e-3}$) results in smaller disks. This difference is mainly caused
by the faster intrinsic growth of the disks for high values of
$\alpha$.  For rather viscous disks, $\alpha = 10^{-2}$, some
difference in the mean size is noticeable near $\sim \SI{500}{au}$,
in the sense that for the simulations without gas the disks are on
average somewhat smaller than in the other simulations. This is caused
by more frequent encounters in the former simulations. The gas
expulsion tends to drive the early evaporation of the cluster, which
leads to larger disks on average because the latter effect terminates
the dynamical disk-truncation process. The mean sizes of the disks in
the simulation with gas but without expulsion tend to be in between the
other two distributions, because some truncation leads to a subsequent
faster viscous growth of the disks, as shown in equation\,\ref{eq:visc}.

The same behaviour can be seen for the average disk masses and accretion rates onto the central star (Fig.\,\ref{fig:CDF}, middle and bottom panel respectively). In the simulations without gas, there are more disks with masses $\lesssim \SI{2}{M_{Jup}}$ than in the other simulations. Again this difference is diminished in the slow viscous evolution case. For the accretion rates the different simulations yield the same final distributions, except for a negligible higher amount of disks with $\dot{M} <\sim \SI{2e-8}{M_{Jup}/yr}$ in the cases without gas. Unlike the case for the average disk radius, using different values of $\alpha$ in the simulations yields comparable results for the final distributions of both average disk masses and accretion rates. The two values for the turbulence parameter used in our simulations yield final values of disk mass and accretion rates that differ by less than an order of magnitude.

\subsection{Evolution of the circumstellar disks}\label{subsec:results_evolution}
\begin{figure}
    \includegraphics[width=\columnwidth]{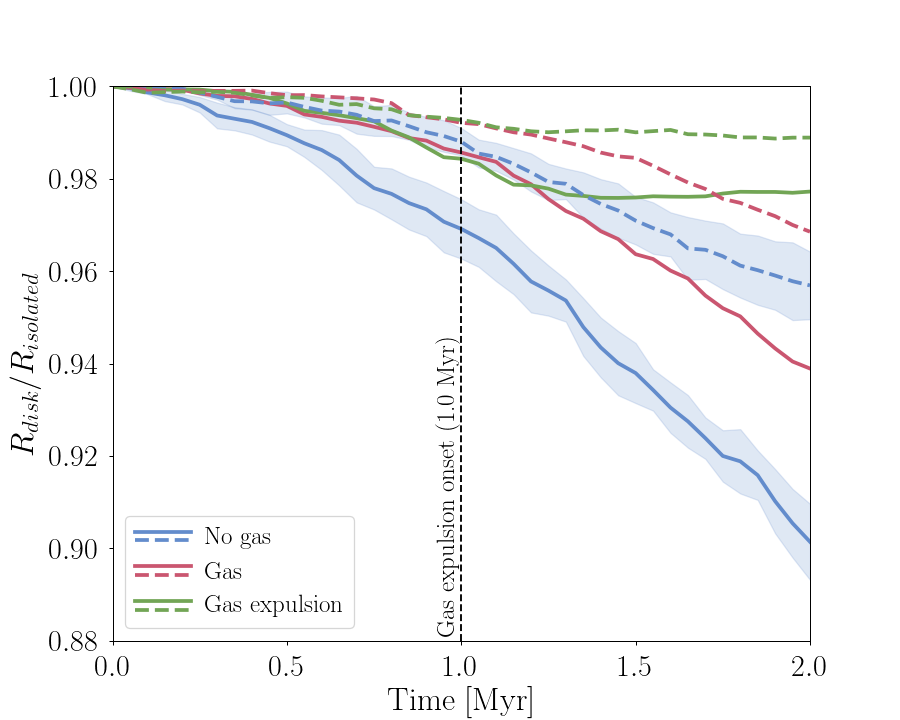}
    \includegraphics[width=\columnwidth]{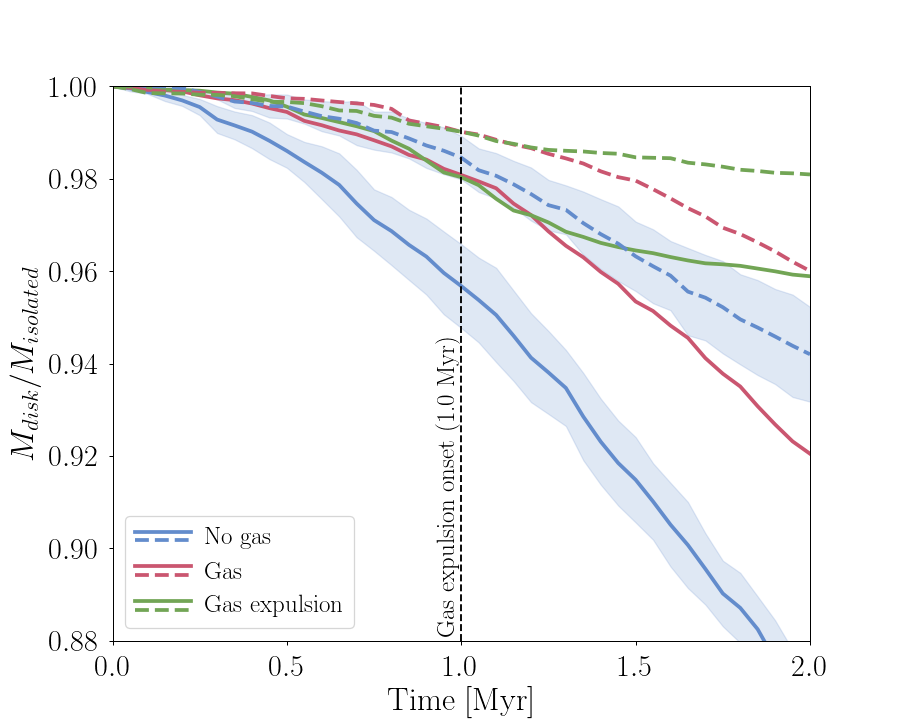}
    \includegraphics[width=\columnwidth]{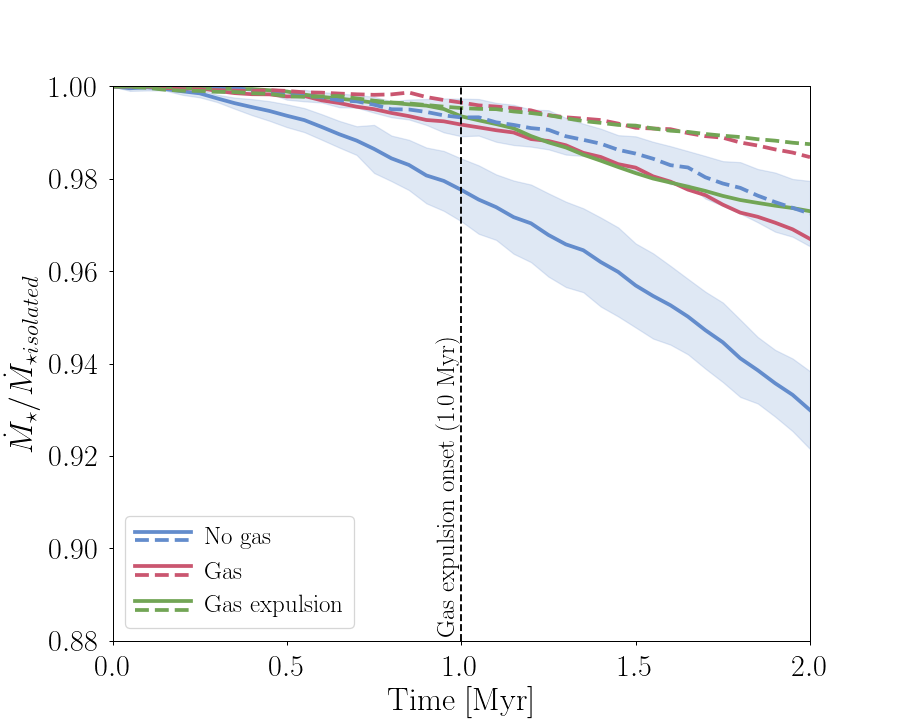}
    \caption{Evolution of the mean disk size (top panel) and mass (center), and accretion rate onto central star (bottom panel) compared to the isolated disk case. By definition, values closer to 1 are similar to the isolated case. The colors indicate the different mode of gas loss in
      the simulation (see legend in the lower left corner). The solid lines correspond to fast viscous disk evolution ($\alpha=\num{e-2}$) and the dashed lines to slow viscous evolution ($\alpha=\num{5e-3}$). The dashed black vertical line shows the gas expulsion onset, $\SI{1.0}{Myr}$. The shaded areas around the blue curves indicate the dispersion around the mean value averaged over 5 simulations. For clarity we only show this uncertainty interval for the blue curves, but the others are comparable.}
    \label{fig:normalized}
\end{figure}

During the $\SI{2.0}{Myr}$ of the simulations, both the viscous growth of the disks and their truncations due to dynamical encounters occur simultaneously. The final disk size, mass, and accretion rate distributions are the result of the combination of these two processes. In Figure \ref{fig:normalized} we show the normalized disk parameters compared to the isolated case, averaged over 5 simulations for each of the gas scenarios. The normalized disk parameters correspond to the actual value for each parameter, divided by the value of the isolated case (viscous growth only, no truncations). By construction, the normalized disk parameters have a value of 1 if there are no dynamical truncations taking place. In a cluster where dynamical truncations are taken into account, we can expect the normalised disk parameters to deviate from 1. The behaviour seen in this case is in agreement with the final parameter distributions shown in Section \ref{subsec:results_gas}. In the top panel of  Figure \ref{fig:normalized} it can be seen how, for the simulations without intracluster gas and for initial disk parameters as specified in Section \ref{subsec:ics}, fast viscous evolution results in disks about 10\% smaller than in the isolated case, while for the slow viscous evolution this value drops to less than 5\%. It can also be seen how the expulsion of the leftover gas allows the disks to simply continue their viscous evolution without being further perturbed by dynamical truncations. These normalized parameters show that the size of the disks in the first Myr of cluster evolution is dominated by the viscous growth, rather than by dynamical truncations. The dynamical encounters experienced by the disks are not enough to truncate them to values largely different from the isolated case. Specially in the case of slow viscous growth, this process seems to be the only one driving the evolution of the disks.

For the disk mass and the accretion rate of the central star (center and bottom panels of Fig.\,\ref{fig:normalized}, respectively), similar behaviours are observed in the curves, also in agreement with the cumulative distributions in Fig.\,\ref{fig:CDF}. Again, the different values of $\alpha$ used in our simulations result in a negligible difference in disk mass and accretion rate. 

\subsection{Comparison with observations}\label{observations}
Observations of circumstellar disks in young clustered environments can help us understand how representative our simulations are. We compared our results with observations of star forming regions and stellar associations. Given the different ages, stellar densities, and general characteristics of the star formation regions mentioned above, we do not look to reproduce precisely their disk distributions using only our approximate model. We carry out this comparison as a way to determine if our model yields reasonable results within the varied collection of young star forming regions.

\subsubsection{Observational data}
We compared our simulation results with observations of star forming regions and stellar associations. The ages, distances, and stellar densities of the observed regions used in this work can be found in Table \ref{data_table}. Given the diverse nature of the observational data used in this work, we give detailed descriptions of the specific observations used for disk sizes, disk masses, and stellar accretion rates. 

\begin{table*}
\caption{Observational values and obtained similar simulations for the observed star forming regions. References: $^{(a)}$\citet{1998ApJ...492..540H}, 
$^{(b)}$\citet{2002ApJ...573..366M}, 
$^{(c)}$\citet{1997AJ....113.1733H}, 
$^{(d)}$\citet{2015ApJ...810...55S}, 
$^{(e)}$\citet{2008hsf2.book..295C}, 
$^{(f)}$\citet{2008ApJS..177..551M}, 
$^{(g)}$\citet{2007ApJS..173..104L}, 
$^{(h)}$\citet{roc2018}, 
$^{(i)}$\citet{1998A&A...332..273B}, 
$^{(j)}$\citet{2017A&A...601A..97S}, 
$^{(k)}$\citet{2004AJ....128.2316S}, 
$^{(l)}$\citet{2016AJ....152..213S}, 
$^{(m)}$\citet{2008MNRAS.383..375C}, 
$^{(n)}$\citet{2006ApJ...651L..49C}, 
$^{(p)}$\citet{2008hsf2.book..235P}, 
$^{(q)}$\citet{2012ApJ...758...31L}}
\centering
\bgroup
\def\arraystretch{1.7}%
\begin{tabular}{cccccc}
                                     & Trapezium  & Lupus clouds                                                                & Chamaeleon I & $\sigma$ Orionis & Upper Scorpio \\
\hline \hline
Age (Myr)                            & $\sim1^{(a)}$    & $1-3^{(e)}$                                                                         & $2-3^{(g)}$          & $3-5^{(k)}$    & $5-11^{(n)}$ \\ 
Distance (pc)                        & $450^{(b)}$        & \begin{tabular}[c]{@{}c@{}}200 (Lupus III)$^{(e)}$\\ 150 (Lupus I, IV)$^{(e)}$\end{tabular} & $\sim190^{(h)}$      & $385^{(l)}$              & $145\pm2^{(p)}$        \\ 
R (pc)                               & $1^{(c)}$          & $\sim52^{(f)}$                                                                    & $4^{(i)}$            & $3^{(m)}$                & $15^{(p)}$            \\ 
N                                    & $\sim2000^{(c)}$ & $\sim12700^{(f)}$                                                                 & $\sim240^{(j)}$    & $340^{(m)}$              & $863^{(q)}$           \\ 
$\rho_N$ (stars pc$^{-3}$) & $\sim250^{(d)}$  & $\sim500$                                                                   & $\sim$0.9    & $\sim$3          & $\sim$0.05    \\ 
Simulation $N$          &  750          & 1000                                                                             &      25        &      25            &   25      \\ 
Simulation $\rho_N$ (stars pc$^{-3}$)     &   1816.09         & 509.35 &    35.25          &       35.25         &         35.25     \\
\begin{tabular}{cc}
    \multirow{3}{*}{Simulation $\alpha$}&Disk size\\
    &Disk mass\\
    &Accretion rate\\
\end{tabular}  & 
  \begin{tabular}[c]{@{}c@{}}$2 \times 10^{-3}$\\ $10^{-4}$\\$10^{-4}$\end{tabular} &
  \begin{tabular}[c]{@{}c@{}}$2 \times 10^{-3}$\\ $10^{-2}$\\$10^{-2}$\end{tabular} &
  \begin{tabular}[c]{@{}c@{}}$10^{-4}$\\ $10^{-4}$\\$10^{-2}$\end{tabular} &
  \begin{tabular}[c]{@{}c@{}}$10^{-4}$\\ $10^{-2}$\\$10^{-2}$\end{tabular} &
  \begin{tabular}[c]{@{}c@{}}$10^{-4}$\\ $10^{-2}$\\$10^{-4}$\end{tabular}
\end{tabular}
\egroup
\label{data_table}
\end{table*}

For the disk radii, we used gas measurements when available. Gas disks are particularly important, because gas dominates the dynamics of the whole disk and gas disks are expected to be larger than dust disks by a factor of $\sim2$ \citep{2018ApJ...859...21A}, since dust can decouple from the gas and drift to the inner regions of the disks. Due to observational constraints, however, gas disks are much more difficult to observe than the compact, sub-mm/mm dusty ones. For this work, we limit ourselves to gaseous radii for disks in the Lupus clouds and Upper Scorpio star forming regions, as noted below.

Given that the chemistry of CO and other tracers of disk mass may be affected by rapid loss of gas or carbon depletion, we chose to use dust masses for our comparisons, scaling them to total disk masses by using the 1:100 dust-to-gas ratio determined by \citet{1978ApJ...224..132B}, which assumes that protoplanetary disks inherit this ratio from the interstellar medium. Recent observations, however, suggest that the dust-to-gas ratio might actually be much lower \citep{2016ApJ...828...46A, 2017A&A...599A.113M}. The implications of this to our analyses are further discussed in section \ref{sec:discussion}.

\paragraph*{Trapezium cluster}
For disk sizes in the Orion Trapezium cluster we used a sample of 135 bright proplyds and 14 disk silhouettes from \citet{Vicente2005}, corresponding to dust radii. The dust mass distribution of circumstellar disks in the Trapezium were obtained from \citet{2009ApJ...694L..36M}. The stellar mass accretion rates were obtained from \citet{2004ApJ...606..952R}.

\paragraph*{Lupus clouds}
Gas radii for 22 circumstellar disks in the Lupus star forming region were obtained from \citet{2018ApJ...859...21A}. Dust masses for 22 disks were obtained from \citet{2017A&A...606A..88T}. Stellar mass accretion rates were obtained from \citet{2014A&A...561A...2A}.

\paragraph*{Chamaeleon I}
Dust radii for 87 circumstellar disks in the Chamaeleon I star forming region were obtained from \citet{2016ApJ...831..125P}. Dust masses for 93 sources were obtained from \citet{2017ApJ...847...31M}. Accretion rates for this region were taken from \citet{2017arXiv170402842M}.

\paragraph*{$\sigma$ Orionis}
Measurements of dust radii and stellar accretion rates for 32 sources in this star forming region were obtained from \citet{2016ApJ...829...38M}. Dust masses for 92 sources were taken from \citet{2017AJ....153..240A}.

\paragraph*{Upper Scorpio}
Gas radii for 57 sources in the Upper Scorpio star forming region were taken from \citet{2017ApJ...851...85B}. Dust masses for the circumstellar disks were obtained from \citep{2016ApJ...827..142B}.

\subsubsection{Preparing simulation results for comparison}
In the previous sections we represented the size of a disk with its characteristic radius, which encloses $\approx63\%$ of its mass (Equation \ref{eq:diskradius}). As a way to do a parallel with actual observations of circumstellar disk sizes, we follow \citet{2017A&A...606A..88T} in fitting the outer radius of a disk at the point where $95\%$ of the mass of the disk is enclosed. To perform the comparisons with observations, we redefined our simulated disk radii as to contain $95\%$ of the disk mass, as follows. Equation \ref{eq:gamma} can be rewritten as

\begin{equation*}
	\Gamma=\left(\frac{R}{\left(1+t/t_v\right)^{1/(2-\gamma)}R_c(0)}\right)^{2-\gamma}.
\end{equation*}
Using Equation \ref{eq:diskradius}, this can be rewritten as

\begin{equation*}
	\Gamma=\left(\frac{R}{R_c(t)}\right)^{2-\gamma}.
\end{equation*}
The radius $R_M$ that encompasses the mass $M$ can be found by solving Equation \ref{eq:mdensity}, from which we obtain

\begin{equation*}
	\left(\frac{R_M}{R_c(t)}\right)^{2-\gamma}=\ln\left(\frac{1}{1-M/M_d(t)}\right).
\end{equation*}
Since we are using $\gamma=1$ in our simulations, this further simplifies to

\begin{equation*}
	R_M=R_c(t)\ln\left(\frac{1}{1-M/M_d(t)}\right)
\end{equation*}
so the radius $R_{0.95}$ that encompasses \SI{95}{\percent} of the disk mass is

\begin{equation*}
	R_{0.95}=\ln(20)R_c(t) \approx \num{3}R_c(t)
\end{equation*}

\subsubsection{Comparison}
The observed star forming regions have distinct ages and stellar densities. This is taken into account when we compare them with simulations. The comparisons were performed as follows: first, the stellar densities of the observed regions were obtained from the literature (observational parameters can be found in Table \ref{data_table}). We performed new simulations with the same initial conditions mentioned in section \ref{subsec:ics}, except now with number of stars $N$=[25, 50, 100, 125, 250, 500, 750, 1000, 1250, 1500, 1750, 2000, 2250, 2500, 5000] and with values of $\alpha$=[$\num{e-4}$, $\num{2e-3}$, $\num{5e-3}$, $\num{7e-3}$, $\num{e-2}$], to allow the observations to be compared with simulated clusters spanning a wider range of parameter space for $N$ and $\alpha$ than the one used for our previous results. Given that the star forming regions with the largest estimated ages are the ones with lowest stellar densities (see Table \ref{data_table}), additional simulations with N=[25, 50, 100, 125] were run for \SI{11}{Myr}. For each observed star forming region, we went through the different simulations and looked at the stellar density at the point in time corresponding to the estimated age for the observed star forming region. The simulation which resulted on the closest stellar density to the one of the observed region was selected as the most similar one. 

For the selected stellar density, there are different values of the turbulence parameter $\alpha$. This means an additional step is needed to determine the simulation result closest to each star forming region. A Kolmogorov–Smirnov (KS) test was performed between the observed and simulated distributions of disk size, disk mass, and accretion rate. A separate KS test was carried out for each of these parameters. 



\begin{figure*}
    \includegraphics[scale=0.55]{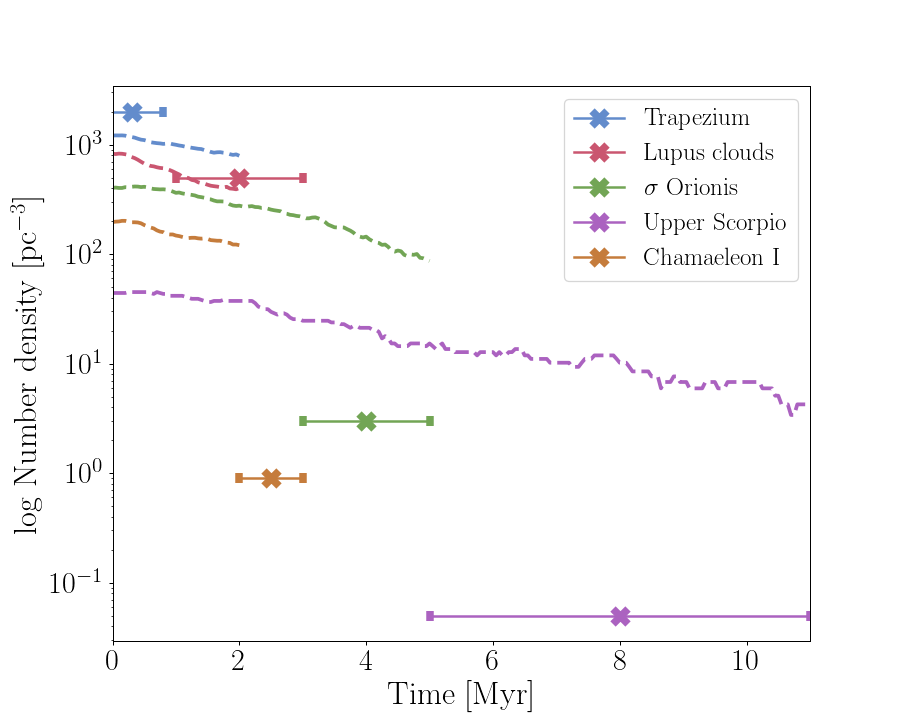}
    \caption{Observed stellar densities of star clusters, along with the most similar simulation. Simulations are plotted at the time comparable to the estimated age of the corresponding cluster (dashed lines in the same colors). The process carried out to find the most similar simulations is described in section \ref{observations}. References for the observational values can be found in Table \ref{data_table}.}
    \label{fig:data_densities}
\end{figure*}

In Figure \ref{fig:data_densities} we show the observational values for stellar density, along with the comparable simulation for each observation. The regions with higher stellar densities (the Trapezium cluster and the Lupus clouds) find good matches in our simulations. For the clusters with lower densities (Chamaeleon I, $\sigma$ Orionis, and Upper Scorpio), even the simulations with the lowest densities are not a good match.

\begin{figure}
    \includegraphics[width=\columnwidth]{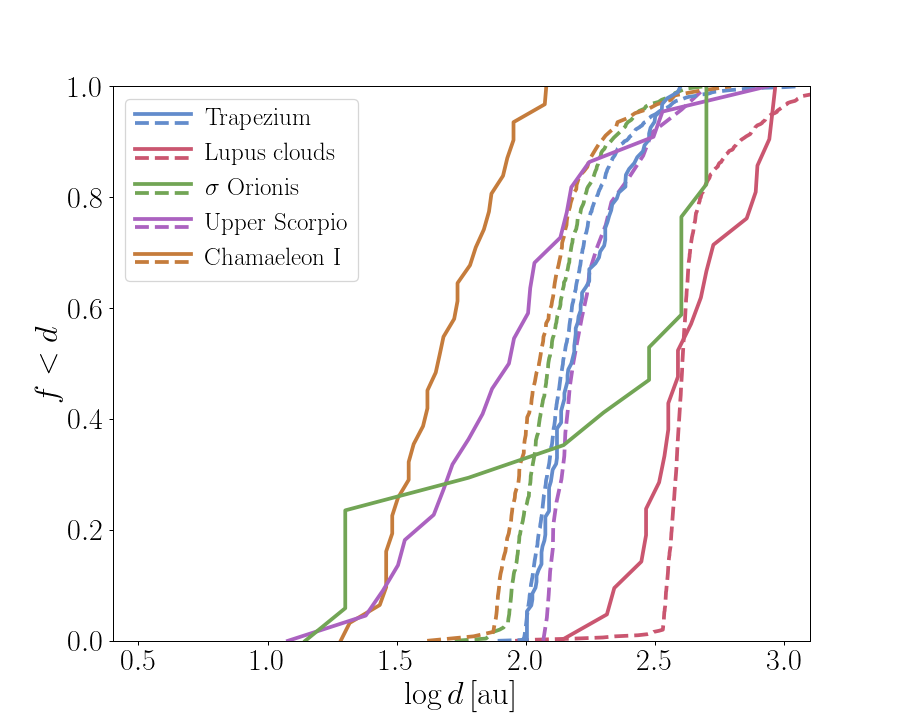}
    \includegraphics[width=\columnwidth]{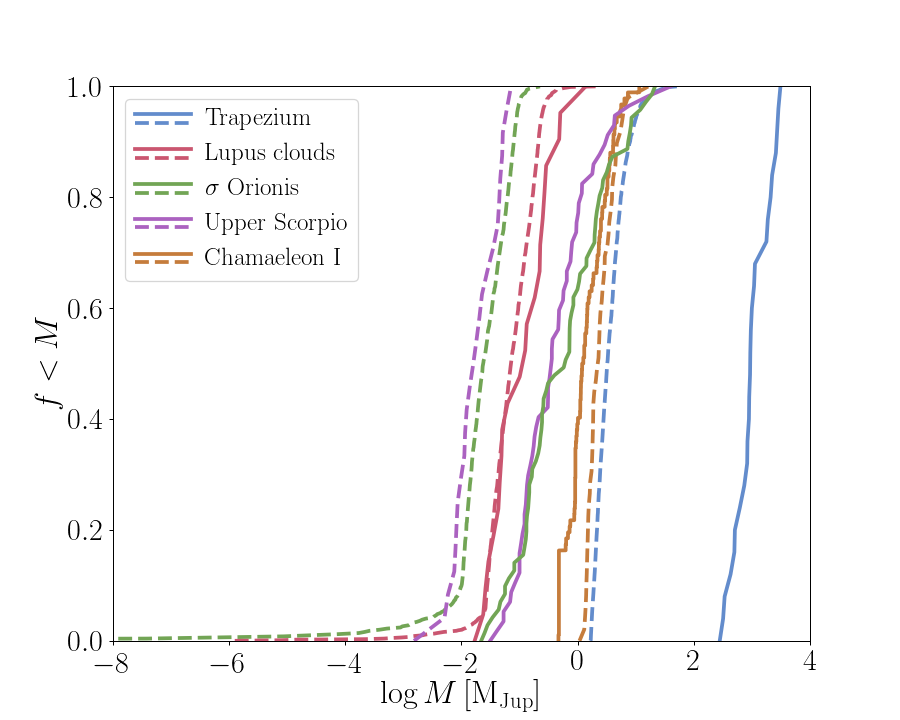}
    \includegraphics[width=\columnwidth]{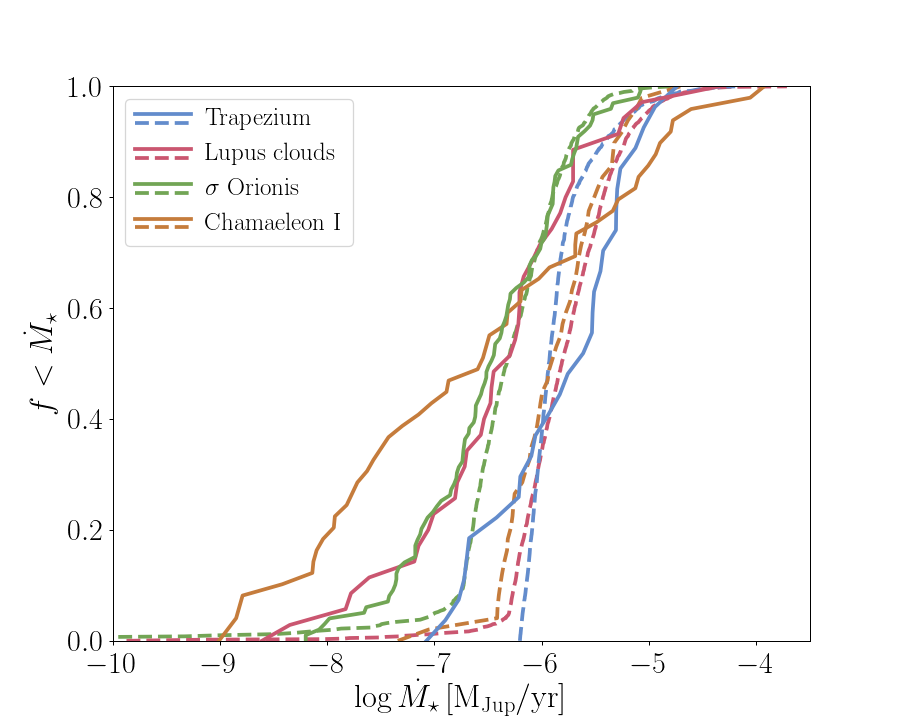}
    \caption{Observed cumulative distributions for the mean disk size (top panel) and mass (center), and accretion rate onto central star (bottom panel) for observed clusters (solid lines) along with the most similar simulation result (dashed lines) obtained at the time comparable to the estimated age of the corresponding cluster. Accretion rates measurements for the disks on the Upper Scorpio region were not available at the time of this paper.}
    \label{fig:cdf_data}
\end{figure}

Given that the results for different gas scenarios in section \ref{subsec:results_evolution} show that disk evolution is dominated by viscous growth rather than dynamical truncations, we are interested in determining if our model is nevertheless able to reproduce the observed distributions of disk sizes, masses, and accretion rates. In Figure \ref{fig:cdf_data} we show the cumulative distributions for disk size, disk mass, and accretion rates for each observed cluster (solid lines) together with its corresponding simulation (dashed line). The simulation curves are plotted at the point in time coinciding with the estimated age of the clusters. The closest resemblance of one of our simulations to an observation is obtained for the Trapezium cluster. The simulation closest to the Lupus region curve both under and over estimates disk sizes. Our model tends to underestimate disk sizes for the top region of the Upper Scorpio observations. This can be related to the fact that gaseous radii where considered for these regions, which leads to large disk sizes \citep{2018ApJ...859...21A}. For Upper Scorpio, however, as reported in \citet{2017ApJ...851...85B} only 4 disks of their sample turn out to have large gas disks. For most of the Upper Scorpio data, as well as for the Chamaeleon I data, our model overestimates disk sizes. 

Regarding the disk masses, in the center panel of Figure \ref{fig:cdf_data} it can be seen that good simulation matches are found for the Lupus clouds and Chamaeleon I. The masses for Upper Scorpio and $\sigma$ Orionis are overestimated by our simulations, whereas the masses for the Trapezium cluster are underestimated.


\section{Discussion}
\label{sec:discussion}

We carried out simulations to understand how the combined effect of viscous disk evolution and leftover gas from the star formation process affect the development of circumstellar disks in star clusters. The disks are subject to viscous growth and can be truncated by dynamical interactions with nearby stars. 

In our simulations we ignore various physical mechanisms that
can alter the size and mass of circumstellar
disks and cluster dynamics over its bound life-time. These effects
include the tidal field of the galaxy, stellar evolution, and
radiative feedback processes.  Initially, our clusters are composed of
single stars each of which has a relatively massive but small ($\sim
30$\,au) disk, in which orientation is ignored and truncation radius is defined as the average over all inclinations.

Photoevaporation of a circumstellar disk can be caused both by the
central star or by nearby OB stars present in the clusters. The
influence of external UV radiation can have an important effect on the outer parts of the
circumstellar disks, causing mass loss and further
diminishing their radii \citep{Scally2001,Guarcello2016}.

Other mechanisms neglected in our model are ram-pressure stripping and face-on accretion on disks.
\citet{Wijnen2016, 2017arXiv170204383W} demonstrated that face-on
accretion of ambient gas in embedded star-forming regions can cause
circumstellar disks to contract, while the ram pressure exerted by the
interstellar medium strips the outer parts of the disks.  Nearby supernovae could also have imporant repercussions on the
morphology and mass of circumstellar disks
\citep{2017MNRAS.469.1117C,2018A&A...616A..85P}, but since our clusters are very young
we ingore this effect. 

Encounters between stars with disks could result in the exchange of
disk-material from one to the other and affect the shape and mass of
both disks \citep{2016MNRAS.457.4218J}.  Such encounters
also tend to harden the surface density of the disks, making their
density profiles diverge from the similarity solutions
\citep{1997MNRAS.287..148H}.


\begin{figure*}
    \includegraphics[scale=0.6]{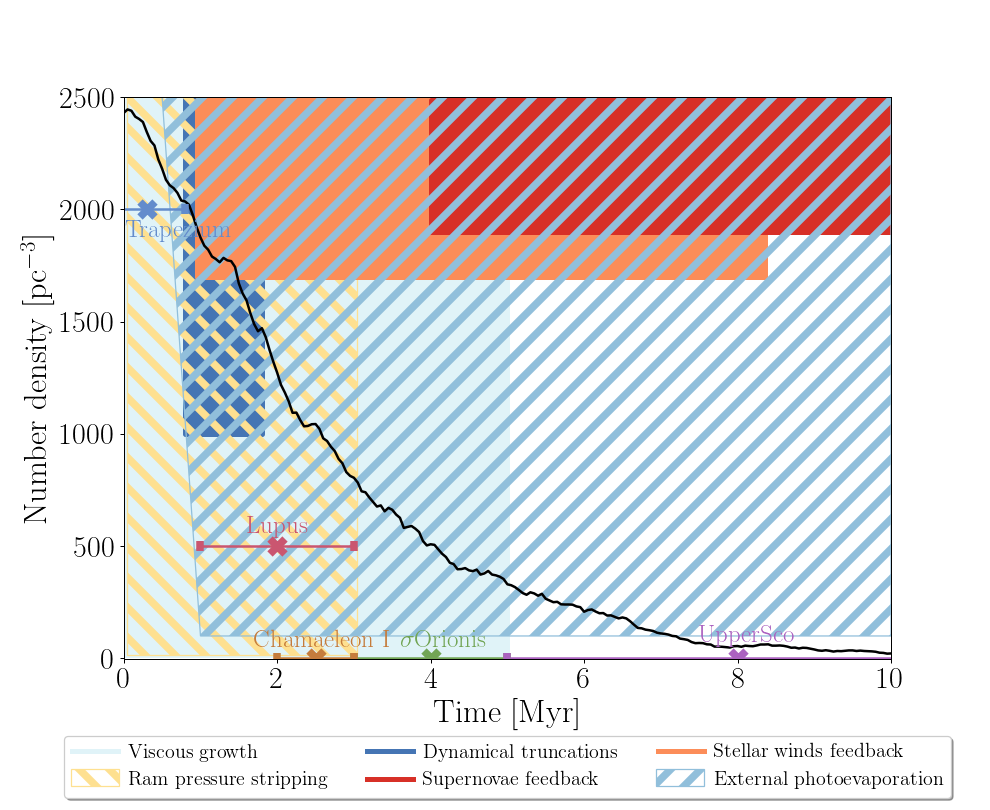}
    \caption{Physical processes in embedded clusters and their corresponding time scales and density scales. The cluster density shown corresponds to one of our simulations for $N=1500$, $\alpha=\num{5e-3}$. References: Ram pressure stripping \citet{Wijnen2016, 2017arXiv170204383W}; Dynamical truncations \citet{PortegiesZwart2016, Vincke2016, 2017arXiv170204383W}; Supernovae feedback \citet{2012MNRAS.420.1503P, 2014MNRAS.437..946P}; Stellar winds feedback \citet{2012MNRAS.420.1503P}; External photoevaporation \citet{2010ARA&A..48...47A, 2013ApJ...774....9A, doi:10.1093/mnras/stw240}.}
    \label{fig:processes}
\end{figure*}

In Figure \ref{fig:processes} we present a schematic overview of the various processes that can affect the final characteristics of circumstellar disks in young star clusters. We take values for viscous growth and dynamical truncations based in our results. We also include ram pressure stripping, stellar feedback from winds and supernovae, and external photoevaporation with values obtained from literature. This figure is intended as a guideline to visualize how incomplete our understanding of the processes that happen inside young star clusters still is. These processes need to be better constrained in parameter space before we can further discuss which ones dominate in the existing observations of young star forming regions. 

Stellar density for a simulation with $N=1500$, $\alpha=\num{5e-3}$ is shown as a guide. Our simulations run only for $\SI{2.0}{Myr}$, and after that the clusters are not dense enough for dynamical truncations to be important. We expand the influence of viscous growth up to $\SI{5.0}{Myr}$. This is the point when the diversity in spectral-energy distributions of observed circumstellar disks settles down and disks are predominantly weak. This means that they could have dissipated or that gas depletion or planet formation could be taking place \citep{2008PhST..130a4024H, 2016SSRv..205..125G}. In our model, as well as in the literature, dynamical truncations do not appear to be a critical process for disk shrinking, in particular when encounters with other stars are distant and when the disks are also affected by external photoevaporation due to bright OB stars \citep{2018MNRAS.473.3223R, 2018MNRAS.tmp..949W, 2018MNRAS.475.2314W}. External photoevaporation can also start in early stages of cluster evolution ($\sim \SI{0.5}{Myr}$) and carry on for almost the whole life of the bright OB stars generating the strong UV radiation \citep{2010ARA&A..48...47A}. Disks can be expected to always be destroyed by external photoevaporation within $\SI{10}{Myr}$ \citep{2013ApJ...774....9A}. Based on the work of \citet{doi:10.1093/mnras/stw240} we extend external photoevaporation down to $N = 100$ stars, since their results show that disks with radius $>\SI{150}{au}$ can endure intense mass loss even for very low ambient far UV fields ($G_0 \sim 30$)\footnote{$G_0 = \num{1.6e-3} \ \text{erg} \ \text{s}^{-1} \ \text{cm}^{-2}$, the interstellar FUV value}. We set the start of external photoevaporation effects at $\SI{1.0}{Myr}$ for low stellar densities, because this is the point where the average disk size for our isolated disks reaches $\SI{150}{au}$. Feedback effects of supernovae start after $\sim \SI{4}{Myr}$. At stellar densities $\gtrsim \SI{1900}{pc^{-3}}$ it can affect the evolution of the disks and even destroy them \citep{2012MNRAS.420.1503P, 2014MNRAS.437..946P}. Winds from nearby stars can affect their neighbors from times as early as $\SI{0.96}{Myr}$; however, stellar densities $\gtrsim \SI{1500}{pc^{-3}}$ are needed for this to affect the evolution of the disks \citep{2012MNRAS.420.1503P}. Ram pressure stripping and face-on accretion affect the disks all through the embedded phase of young star clusters, even for very low stellar and gas densities \citep{Wijnen2016, 2017arXiv170204383W}.

The discrepancy between observations and simulation results could reflect not only the need to include more physical processes. The initial conditions chosen for the simulations could also not be representative of real clusters. It is possible that young stellar clusters present substructure, in which the local stellar density might be higher and in turn lead to more dynamical encounters \citep{2004A&A...413..929G}.

We overlook the presence of primordial binaries in our initial
conditions, which could also contribute to the overestimation of our 
disk sizes. \citet{2017arXiv171103974C} show that disks around binary
stars tend to be smaller than their isolated counterparts, and they
tend to be less bright. Neglecting primordial binaries is a rather
strong assumption, because they tend to have a strong effect of the
disk-size distribution and their survivability in the cluster. We
realize that our simulations tend to overestimate disk sizes, but
observations may as well underestimate disk sizes, in particular of the outer
extended regions of disks where they have a low surface density.

The different descriptions of observed disk radii and masses used together in section \ref{observations} may also explain why we do not see consistency in the over and underestimation of these parameters in our simulations. Having a uniform description of the observational data would be ideal to perform a more accurate comparison. Thanks to Gaia DR2 \citep{2018A&A...616A...1G}, the distances to the star forming regions considered in this work are being calculated more precisely \citep[e.g.][]{roc2018}, which could also reflect differences in disk sizes from the ones here reported. The first-order approach obtained in this paper, however, serves as a good guideline as to where to direct future developments.

\section{Summary and conclusions}
\label{sec:summary}

We studied the effect of viscous growth and dynamical truncations of circumstellar disks inside young star clusters. We used a semi-analytic model to include the viscous evolution of the disks, and a background potential to implement the gas in the cluster. We studied three scenarios for the gas: no gas in the cluster, constant gas through the cluster's evolution, and gas expulsion halfway through the cluster's evolution. For this, we ran simulations with number of stars $N=1500$, turbulence parameter $\alpha=\num{e-2}$ and $\alpha=\num{5e-3}$, and spanning $\SI{2.0}{Myr}$ of cluster evolution.

Our simulations result in similar distributions for average disk size, disk mass, and accretion rate onto the central star, independent of the gas in the cluster. Although clusters without leftover gas result in a higher amount of disks with radius $\lesssim \SI{500}{au}$, in our simulations gas presence does not seem to largely shape the final distribution of circumstellar disk parameters. In an environment where dynamical truncations were important in shaping the sizes of the circumstellar disks, we would expect the presence of gas to make a difference in the final disk size distribution, since stars still embedded in leftover gas have higher velocity dispersions \citep{Vincke2016} which in turn leads to more dynamical encounters. In the parameter space spanned by our simulations, dynamical truncations are overtaken by viscous growth in determining the size of the circumstellar disks. 

The different values of the turbulence parameter are reflected in the resulting sizes of circumstellar disks. Simulations with fast viscous growth ($\alpha=\num{e-2}$) return bigger disks, but these disks are still not large enough to be affected by dynamical encounters. The size of the circumstellar disks in our simulations is only defined by the inherent viscous growth of the disks. Dynamical truncations do not play an important role in the determination of the final disk sizes and masses.

We performed a comparison of our simulation results with observational data of circumstellar disk sizes, masses, and stellar accretion rates in several young star forming regions. To better match the stellar densities of the observed regions, we expanded the parameter space of our simulations to number of stars $N$=[25, 50, 100, 125, 250, 500, 750, 1000, 1250, 1500, 1750, 2000, 2250, 2500, 5000] and values of $\alpha$=[$\num{e-4}$, $\num{2e-3}$, $\num{5e-3}$, $\num{7e-3}$, $\num{e-2}$]. We also adjusted the definitions of size and mass of our simulated disks to suit the descriptions of the observed disks.

Low values of the turbulence parameter ($\alpha$=$\num{e-4}$) are not enough to reproduce the small circumstellar disk sizes observed in the star forming regions. Simulations with higher values of $\alpha$ ($\alpha$=$\num{5e-3}$ and higher) differ even more from the observed disk distributions. The stellar density of the simulated clusters is not enough for dynamical encounters to actively truncate the disks and reproduce observed circumstellar disk sizes. Dynamical truncations by themselves are not relevant enough to shape the observed distributions of circumstellar disk sizes and masses. Other processes are at play in terms of counteracting the viscous growth of the disks.

Compared to observations, our model both under and overestimates different disk parameters, but does not show a consistent behaviour related to the data. This could be due to the physical processes ignored in this work, or to an incorrect selection of initial conditions. It is also important to note that the observational data is not uniformly characterized, which could contribute to the incongruency with simulation results. The distributions of disk parameters obtained by our simulations, if not accurate, still fall within ranges in agreement with the ones spanned by observations of different star forming regions.

\section*{Acknowledgements}

We thank Maxwell Cai, Thomas Wijnen, Ana Miotello, Michiel Hogerheijde and the protoplanetary disk group at Leiden Observatory for valuable
discussions. We also thank the anonymous referee for their comments which helped make this paper better.
This work was supported by the Netherlands Research
School for Astronomy (NOVA) and NWO (grant \#621.016.701 [LGM-II]). This paper makes use of the packages \texttt{numpy} \citep{numpy:2011}, \texttt{scipy} \citep{scipy:2001}, \texttt{matplotlib} \citep{Hunter:2007}, and \texttt{makecite} \citep{makecite:2018}.


\bibliographystyle{mnras}
\bibliography{tex/references}

\bsp	
\label{lastpage}
\end{document}